\documentclass[fleqn,twoside]{article}
\usepackage{espcrc2}

\usepackage{graphicx}
\usepackage[figuresright]{rotating}

\newcommand{\AmS}{{\protect\the\textfont2
  A\kern-.1667em\lower.5ex\hbox{M}\kern-.125emS}}

\title{Cosmic Ray Positrons from a Local, Middle-Aged Supernova Remnant}

\author{$^{1}$A.D.Erlykin\address{P.N.Lebedev Physical Institute, Moscow, Russia},
        A.W.Wolfendale\address{Department of Physics, Durham University, Durham, UK}}%

\begin{document}

\begin{abstract}
We argue that the cosmic ray positron excess observed in ATIC-2, Fermi LAT, PAMELA, 
HESS and recently in the precision AMS-02 experiment can be attributed to 
production in a local, middle-aged supernova remnant (SNR). Using the 
prediction of our model of cosmic ray acceleration in SNR we estimate that the SNR 
responsible for the observed positron excess is located between 250 and 320pc from 
the Sun and is 170-380 kyear old. The most probable candidate for such a source is the 
SNR which gave birth to the well-known Geminga pulsar, but is no longer visible. Other 
contenders are also discussed.   
\end{abstract}

\maketitle

\section{Introduction}
\footnote{Corresponding author: tel +74991358737 \\ 
 E-mail address: erlykin@sci.lebedev.ru}
            Countless papers have presented evidence for cosmic ray (CR) 
nuclei and electrons below some PeV energies, at least, as having been 
accelerated in Supernova Remnants (SNR).
            We, ourselves, have gone further and made the case for the 
well known 'knee' in the CR energy spectrum being due to a single, 
nearby, recent SNR (see \cite{EW1} and later papers.). 
Similar sources should have specific implications for the minority 
electron component. Indeed, because of their higher energy losses, 
nearby sources should give a bigger fraction of the measured flux and 
spectral structure should result. Such structure has, in fact, been 
claimed (\cite{Panov}) although specific sources have not yet been 
identified and pulsars as well as SNR have been put forward as the sources.

    Positrons are another minority component and interesting observations 
of the excess positron flux over expectation have been made by the 
PAMELA and Fermi LAT instruments \cite{Adria,Acker}. Very recently, 
the AMS-02 instrument has confirmed the earlier results \cite{Aguil}

 The origin of the increasing positron fraction has been discussed in many 
publications and the majority follow the view of a single, nearby 
source, most likely a pulsar, being responsible, although an annihilation of 
 dark matter particles cannot be excluded. Here, we put forward an alternative 
view: that the single source is, in fact, an SNR and not a pulsar. 
We start by examining the arguments in favour of a SNR as the source
of positrons rather than a pulsar.
\section{Why SNR ?}
Our examination is founded on the results of AMS-02 data \cite{Aguil}
as being the most precise. The AMS-02 collaboration presented their results
in terms of the positron fraction $\Phi$ of the total flux of electrons and 
positrons $\frac{e^+}{e^+ + e^-}$ as a function of energy $E$. This 
fraction is shown in Figure 1 by open circles. 
\begin{figure}[htb]
\begin{center}
\includegraphics[height=8cm,width=7cm,angle=-90]{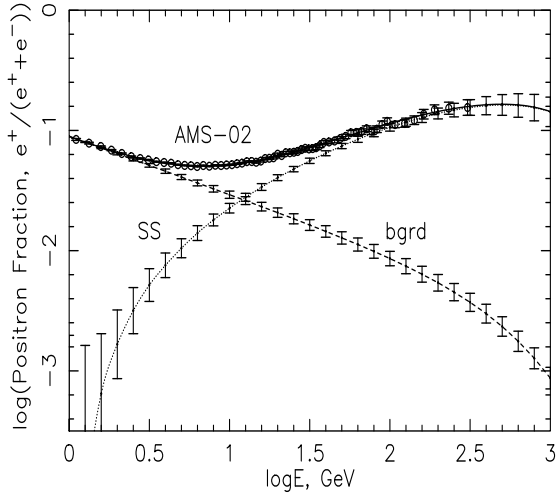}
\end{center}
\caption{\footnotesize Formation of the positron fraction as a function of 
energy in the AMS-02 experiment. Open circles - experiment, fitted by a thin 
full line according to expression (1). This fit is extrapolated to energies 
above 350 GeV up to 1000 GeV. Errors of the experimental data are obtained by 
the summation of squared statistical and systematic errors. The fraction is divided into 
 two parts: $e^+$ from the background of a variety of sources - dashed line 
denoted as $bgrd$, and from the single dominant source - dotted line denoted as $SS$. 
Errors of the background fraction are taken as due to an uncertainty in the slope index
 equal to 0.03. Errors of the SS-fraction of positrons from the single source 
are calculated as the sum of the squared errors of the experimental data and errors of 
the subtracted background.}   
\label{fig:fig1}
\end{figure}

The collaboration analysed their experimental data in terms of the so called 'minimal 
model', where the electron and positron energy spectra were described  by the 
expression (1):
\begin{equation}
\Phi_{e^{\pm}} = C_{e^{\pm}}E^{-\gamma_{e^{\pm}}} + C_sE^{-\gamma_s}exp(-E/E_{cut})
\end{equation}
The first term in this expression describes the contribution of the diffuse
power law spectra of electrons and positrons and the second one relates to the 
contribution of the dominant local source. The former term for electrons relates to the
 background  of 'primary' electrons accelerated by the variety of sources in the Galaxy
 and propagated to the Solar system. The background of positrons relates to 
secondaries from interactions of the majority CR component with gas nuclei in the 
Interstellar Medium (ISM).

 In the 'minimal model' adopted by the AMS-02 collaboration the latter term of 
expression (1) contributes equally to the electron and positron flux which may be 
inspired by the possible pulsar origin of the positron excess. However, this model is 
not unique. The experimental data could be also fitted with the functions in which the 
local source emits non-equal amounts of electrons and positrons. For example, even the 
extreme assumption that the single source emits only positrons gives an equally good 
fit with a slightly modified constant (~$C_s/C_{e^-}$ = 0.0070 instead of 0.0078 in 
\cite{Aguil}~). In the following analysis of the characteristics of the positron 
fraction we used the fit suggested by the AMS-02 collaboration but keep in mind that it
does not mean that the real single source emits equal amounts of electrons and 
positrons. Therefore it needs not necessarily be the pulsar.

In the framework of the leaky box model the diffuse background spectrum of positrons 
should be 
steeper than that of electrons with the difference $\delta$ of the slope indices equal 
to that of the energy dependence of the CR lifetime in the Galaxy \cite{Blasi}. AMS-02 
measurements give $\gamma_{e^+} - \gamma_{e^-} = 0.63\pm 0.03$, which agrees well with
the estimates of $\delta$ obtained from the measurements of the ratio of secondary to 
primary CR nuclei \cite{Gupta}. 

Within the leaky box model the same index $\delta$ describes the difference between the
 CR energy spectra injected from the sources and those observed after their 
propagation in the Galaxy. The CR proton spectrum observed during the previous AMS-01 
flight 
had the slope $\gamma_P = 2.78 \pm 0.009(stat) \pm 0.019(sys)$ \cite{AMS-01}. If the 
difference between the slopes of the observed and injected spectra is $'\delta'$ which 
is equal to $0.63 \pm 0.03$, then the injected spectrum should have 
$\gamma_{inj} \approx 2.15 \pm 0.04$.
This value coincides with that expected by us for the emergent spectra injected from 
SNR \cite{EW2} (~although it is appreciated that a variety of factors undoubtedly give 
rise to a range of slopes for the injected CR~). The injected spectra of electrons from
the SNR in our model are the same as that of protons, ie they have 
$\gamma_{inj} \approx 2.15$. Positrons as the secondary particles originate from the 
interactions of CR with ISM after they propagate in the Galaxy, therefore their 
'injected' spectrum has the slope index equal to $\gamma_{inj} + \delta$.    

This  consideration relates to the so-called 'background' spectra of electrons and 
positrons. The coincidence between the difference of slope indices with that between
slopes of observed and injected ('source') spectra of protons accelerated in SNR 
indicates that the source of background electrons and positrons are SNR. We assumed 
that the source of positron excess observed at high energies in many experiments 
including AMS-02 can be also the SNR which accidentally occured in the recent time and 
nearby the Solar system.    

We start by examining the case that can be made for 'primary' positrons being 
accelerated by SNR shocks and this is followed by a derivation of the distance and age 
of the SNR that could be responsible. The model adopted is that advocated consistently 
by us, and referred to above.
\section{Positrons from an SNR}
\subsection{The mechanism}
           The idea of 'diffusive shock acceleration of decay positrons 
in SNR' is not a new one. Ellison et al. made the suggestion , with particular emphasis
 on positrons of energy below 10 MeV, the positrons escaping and annihilating with 
electrons and  thereby generating gamma rays of energy 0.511 MeV \cite{Ellis}. More 
recently, Zirakashvili and Aharonian have applied the mechanism to higher energy 
positrons \cite{Zirak}. Here, we consider the acceleration of positrons to some 
100s of GeV.
\subsection{Sources of positrons in SN ejecta}
   A number of radioactive nuclei from SN ejecta are positron emitters, 
principally: $^{26}$Al, $^{44}$Ti, $^{56}$Co, $^{56}$Ni and $^{57}$Ni. The mean 
positron energy is about 1 MeV and is thus 'high' in comparison with the  
thermal energy of (negative) electrons. The higher energy allows the 
positrons to be injected into shocks with high efficiency. (~Dieckmann  
et al. argue that $10^5$ eV is the threshold energy for electrons for 
such injection, \cite{Dieck}~). The preference for positrons to be injected is reduced 
somewhat, however, by the fact that in the SNR there will be many more 
pre-existing electrons from the (ISM) and these are potentially 
available for acceleration.
 
  It is well known that the number of available positrons is very 
variable, depending, as it does, on the SN type, and precursor stellar 
mass. Chan and Lingenfelter have examined the problem in detail \cite{Chan}. 
 Their results give survival fractions up to 30\%, although many models 
have much smaller values. Presumably, the escape probability from the 
dense initial SN environment will be low for positrons produced 
only a few days after the SN explosion; those from $^{56}$Co are a case in 
point, for this nucleus the output is large (approaching a stellar mass 
; \cite{Chan}) but the half life is short; 77 days. 
Nevertheless, if the magnetic field is in the 'combed mode', ie 
streaming away from the SN, the problem will be eased. It is true that radio studies
give little evidence for such streaming modes but it will be realised that the 
particular SNR hypothesised to generate the detected positrons has disappeared by now; 
thus the combed mode must be regarded as an assumption for this particular SNR.

   We consider that a reasonable case can be made for the upturn in the 
positron spectrum being due to a local SNR and in what follows 
we use our 'standard model' to evaluate the predicted distance and age 
of this SNR.
\subsection{Checks on the hypothesis}
Although we favour a SNR as being responsible for the extra positrons, others prefer a 
pulsar. Our preference can be justified by two observations. \\
(i) The source energy spectrum of CR emitted by the pulsar is expected to be much 
flatter than that from SNR. Many authors ( for example, see \cite{Ostr,Blas1,Gill} ), 
including ourselves \cite{EW3}, conclude that the slope index of the emergent spectrum 
has to be as small as $\gamma_s = 1$. It is much less than the $\gamma_s \approx 2.15$
expected from AMS-02 data for single source positrons. Also, if the pulsar is nearby 
and young, the 
propagation effects for its CR have to be small and the expected spectrum of positrons
should also be much flatter than observed. \\ 
(ii) Another distinction  would be expected by way of the presumption that the SNR 
would be a source of extra positrons only ( their origin being via the radioactive 
decay of the SN ejecta ) whereas the pulsar would be a source of equal numbers of 
electrons and positrons. Future publications ( particularly from AMS-02 ) in which 
electrons and positrons will be distinguished and their energy spectra have a good 
precision will show an electron spectrum with an upturn in intensity starting at a 
little below 100 GeV (~as for positrons~) if a pulsar is responsible but perhaps 
a smaller one if an SNR is involved with its emphasis on positrons. However, it must 
be remembered that a local source will also give a contribution from ambient electrons 
accelerated in the SNR itself (~see, for example \cite{EW4}~). Measurements of $e^+$ 
and $e^-$ at higher energies than at present are crucial in this regard; perhaps only 
if the positron fraction exceeds 50\% will it be possible to conclude that positrons 
from SNR ejecta predominate.

The points favouring the SNR origin of the positron upturn must be qualified by the 
necessary assumptions: that the SNR magnetic field must have been in the 'combed mode'
and that our SNR acceleration model is applicable to the SNR positrons. In view of the 
disappearance of the SNR itself direct studies of the actual environment in which the 
SN exploded are difficult. However, subtleties may remain which favour the important 
assumption about the mode of the magnetic field lines (~which helped the positrons 
escape~).

\section{The derivation of the positron energy spectrum from the single source}
In our analysis we follow the scenario proposed by the AMS-02 group. They assume that 
the positron spectrum is the sum of two parts: from the background and from the single 
source. These two parts are described respectively by the left and right terms of the 
expression (1). The left background term is a simple power law. We assumed that the 
contribution of the single source at the lowest energies is negligibly small and 
determined the power index of the background spectrum at these GeV energies from the 
best fit of the positron fraction measured by AMS-02. The result was 
$\gamma_{e^+} = 3.48 \pm 0.03$. The accuracy of 0.03 was taken from that of 
$\gamma_{e^-} - \gamma_{e^+}$ given by AMS-02. 

The best fit of AMS-02 data requires $\gamma_{e^-} - \gamma_s = 0.66 \pm 0.05$ and 
$\gamma_{e^+} - \gamma_{e^-} = 0.63 \pm 0.03$ \cite{Aguil}. Hence 
$\gamma_{e^+} - \gamma_s = 1.29 \pm 0.06$ and subtracting the last expression from 
$\gamma_{e^+} = 3.48 \pm 0.03$ one can obtain $\gamma_s = 2.19 \pm 0.06$ which agrees 
well with the value of 2.15 for the slope of the emergent spectrum from SNR \cite{EW2}.
It gives more support to our assumption that the source of positrons can be an SNR and
that energy losses en route are small. The single source spectral shape differs from 
that for the ambient electron spectrum of course, which has in it the Galactic loss 
parameter with the exponent of the energy dependence equal to 0.66.  

We extrapolated the power law spectrum with the slope index of $\gamma_{e^+} = 3.48$ to
 higher energies. It is shown by the dashed line in Figure 1. Since numerous 
exprimental data and simulations indicate that at high energies approaching the TeV 
region the electron energy spectra have a cutoff due to the rising energy losses we 
applied the same cutoff term of $exp(-E/E_{cut})$ to our diffusive background spectrum.
$E_{cut}$ has been taken equal to 760 GeV - the value adopted by the AMS-02 
collaboration with which we agree. It has a small steepening effect at energies close 
to 1 TeV as seen in Figure 1, but this has a negligible effect on the predicted 
magnitude of the single source spectrum.
    
The contribution of the single source is obtained by subtraction of the background
from the positron fraction measured by AMS-02. It is shown by the dotted line in Figure
 1. In what follows we will analyse this single source spectrum to derive the possible
distance and the age of the assumed SNR responsible for its formation.
\subsection{The distance and age of the SNR responsible for the positron excess}
In \cite{EW4} we described our simulation program used to analyse the CR electron 
spectra. It was also used here for the analysis of the positron spectrum from the 
single source described in the previous subsection. It is assumed that the emergent 
positron spectrum has the same spectral shape as that of the electrons accelerated in 
'conventional' SNR. This positron spectrum was obtained as the product of the total 
electron and positron spectrum and the fraction
of positrons from the single source. 

Our simulation program did not distinguish electrons and positrons so that we consider 
its output as the total spectrum of electrons and positrons. The detailed description 
of the program can be found in \cite{EW4} (~a relevant feature was the adoption of 
'anomalous diffusion' for the propagation of CR from source to Earth~). Here, it is 
enough to say that we simulated 
50 different spectra obtained by the summation of contributions from 50000 SNR 
randomly distributed in the local part of our Galaxy with the radius $R < 3.16 kpc$ 
centered on the Sun and in a time range up to $10^8$ years. The variety of spectra due 
to the different samples of SNR distributions in space and time can be seen in Figure 
4a of \cite{EW4}. Here, we use for the total electron and positron spectrum the median 
of the 50 simulated spectra.

We multipled this total electron and positron spectrum by the fraction of positrons 
from the single source obtained from the analysis of AMS-02 data described in the 
previous subsection and obtained the absolute spectrum of positrons from the single 
source. Both are shown in Figure 2 by thick lines: dashed - for the total $e^+ + e^-$ 
spectrum, full - for the expected single source $e^+$ spectrum.
\begin{figure}[hpt]
\begin{center}
\includegraphics[height=15cm,width=7cm]{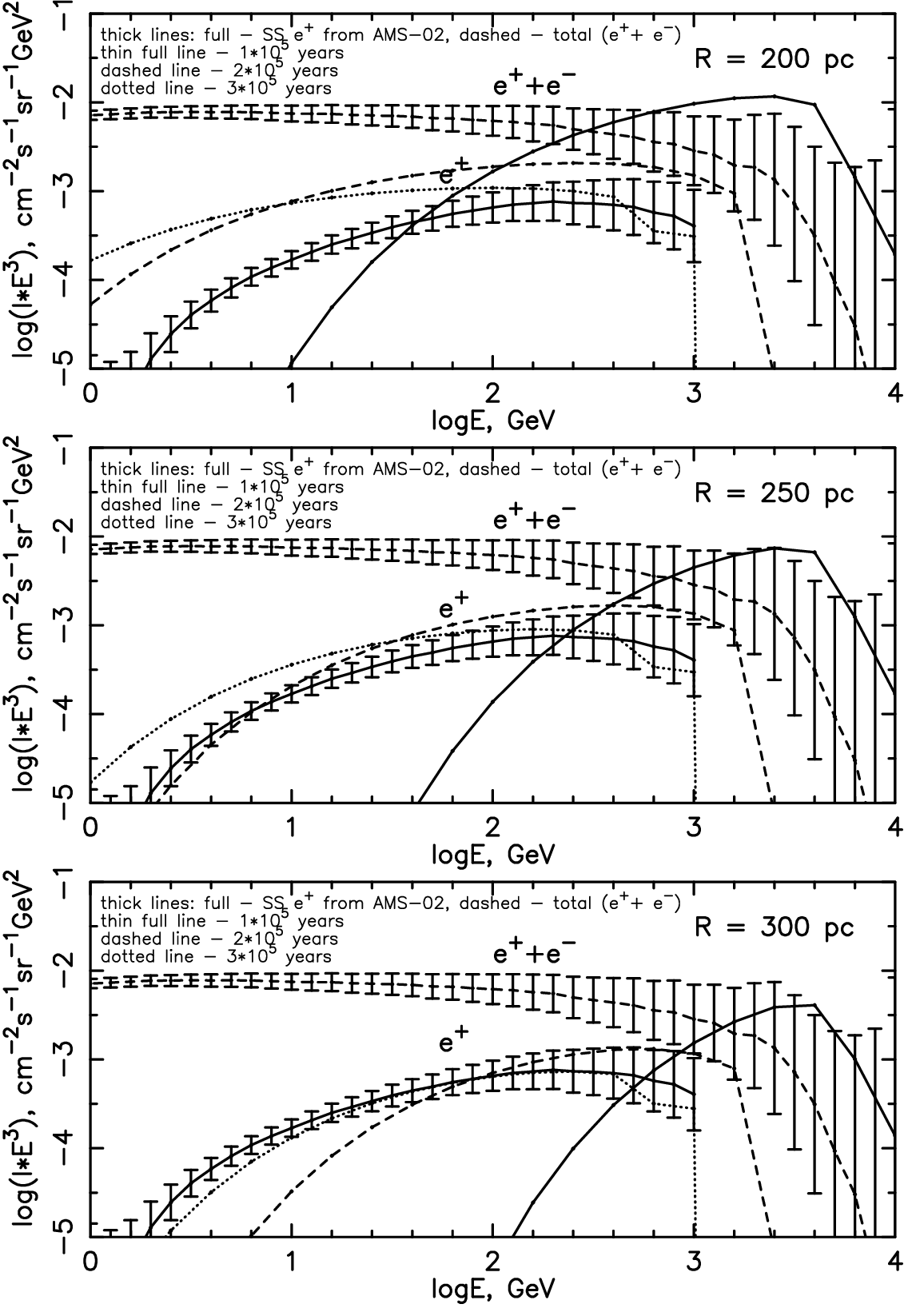}
\end{center}
\caption{\footnotesize Examples of the positron spectra calculated for the SNR of 
different ages: 100 (thin full line), 200 (dashed line), 300 kyear (dotted line) and
 located at different distances: 200, 250 and 300 pc from the Sun (upper, middle and 
lower panels respectively). The total $e^+ + e^-$ spectrum and single SNR $e^+$ 
spectrum derived from the AMS-02 data are shown by thick dashed and full lines,  
respectively.}   
\label{fig:fig2}
\end{figure}   

The same program was used to calculate positron spectra from SNR of different ages and 
of different distances from the Sun. The calculations were made in the age interval of 
180-400 kyear with a 10 kyear bin and within the distance range of 230-350 pc with 
a 10 pc bin. Some results of such calculations are shown in Figure 2. Calculated 
spectra were compared with that obtained from the AMS-02 data using the $\chi^2$ test.
No normalization has been applied because both experimental and calculated spectra were
obtained using the same simulation program \cite{EW3}. Comparison has been made in the 
energy interval from 10 to 1000 GeV since at lower energies the experimental data are
distorted by solar modulation effects which are not taken into account in the 
simulation program. 

The contour plot in the age-distance diagram  drawn through the points where 
$\chi^2/ndf = 2$ is shown in Figure 3. 
\begin{figure}[ht]
\begin{center}
\includegraphics[height=8cm,width=6cm,angle=-90]{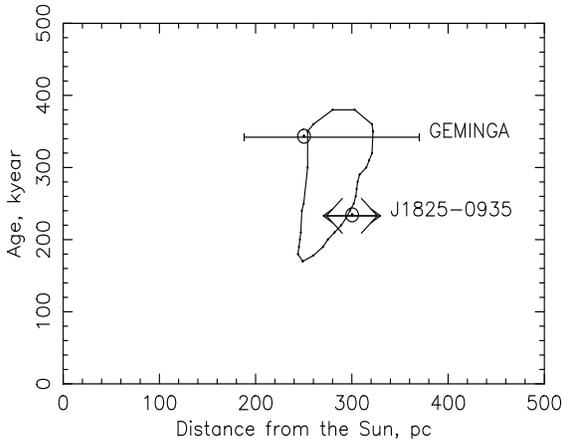}
\end{center}
\caption{\footnotesize The age-distance diagram for the determination of the possible
source of positrons in the AMS-02 experiment. The area covered by the contour shows the
 range of ages and distances from the Sun where the comparison with the experimental 
data with the model gives $\chi^2/ndf < 2$. Open circles show the positions of the 
Geminga and J1825-0935 pulsars which can trace the location of the SNR responsible for 
the positron excess. The age indicated is the so called 'characteristic age': 
{\it P/\.P}, where {\it P} and {\it \.P} are pulsar period and its derivative, 
respectively. Since they are usually determined with a high precision the uncertainty 
of the 'characteristics age' is negligibly small.} 
\label{fig:fig3}
\end{figure}
\section{Identification of the SNR satisfying the age and distance requirements}
We searched for a possible candidate to be a source of the positron excess among those 
which could hit the region of the age-distance plot indicated in Figure 3. If the total
 SN explosion rate in the Galaxy with radius of 15 kpc is about 0.02 y$^{-1}$ then 
the expected mean number of SNR which satisfy the obtained selection requirement
is about 0.3 and it is unlikely that there are more than one or two sources within this
 region. 

Since it is likely that all pulsars are produced in SN explosions and that the 
well-known lack of SNR associations is simply a problem of age, ie 'old' pulsars have 
'lost' their SNR, we looked for the possible candidate both through SNR and pulsar 
catalogs. We have found only one 'good' candidate - the Geminga pulsar. Its distance
determined by the parallax measurement is 250+120-62 pc \cite{Faher} and its age is 330
 kyear. Its position on the age-distance plot is shown in Figure 3. The closest 
approach of calculations to the derived energy spectrum of positrons from the single 
source is indicated by the dotted line in the middle panel of Figure 2.  

There is another pulsar, J1825-0935, which is 233 kyear old. In the latest version of 
the ATNF catalog it is located at 300 pc from the Sun, but the uncertainty of this 
distance is very high because in the previous versions of the catalog this value spans 
the range from 
0 to 2400 pc.

It is necessary to point out that the estimated ages are most likely lower limits. 
They are usually calculated using a braking index of 3. In most cases when it is 
possible to measure this index, however, it appears to be less than 3 and the 
corresponding age should be higher.

It is worth mentioning that the famous Loop I SNR cannot be excluded as the possible 
candidate. It is not shown in Figure 3 because the uncertainty of its distance and age
are extremely high. 

We now discuss in more detail the properties of Geminga as the most 
probable single source of the positron excess.   
\section{Geminga}
The Geminga pulsar and its pulsar wind nebula (PWN) were already examined as the 
possible source of the observed positron excess (~see \cite{Yuks} and references 
therein~). It is necessary to examine here a number of facts about the Geminga pulsar: 
its distance, age and the possibility of there having been an associated, parent, SN, 
the remnant from which has since been 'disssolved' into the general ISM. 
\subsection{Distance and age}
A number of studies have been made which give the distance and/or age and there have 
been summarised to give the limits shown in Figure 3. We are mindful of the fact that 
there is an error due to the pulsar's proper motion since it was formed in the SN 
explosion: this is 
$\sim$30 pc for a typical velocity of 100 km$\cdot s^{-1}$. It is neglected in 
comparison with the other uncertainties. The data adopted come from those already 
mentioned \cite{Faher} and also from the works \cite{Gehr,Cuhna,Carav,Danil,Brazi}.   
\subsection{An invisible SNR associated with the Geminga pulsar as the possible 
positron source}
Geminga was discovered as a gamma ray source without an associated SNR or radio halo.
It is assumed that due to its relatively great age the pulsar lost its SNR. The 
question 
is a quantitative one - at what age do the SNR become non-identifiable ? We have 
studied the pulsar catalog \cite{Lyne} to answer this question. Confining attention to 
pulsars within 3 kpc, loss appears to start at 20 kyear and reaches about 90\% by an 
age of 300 kyear, the age of Geminga. Thus, the lack of an observed SNR associated with
 Geminga is quite understandable.

For Geminga itself, it has been suggested \cite{Cuhna} that the expanding ring of gas 
surrounding Lambda Ori could be due to a SNR explosion that occured 300-370 kyear ago.
We conclude that there is a good case that there was a Geminga SNR.
\subsection{Relevant characteristics of Geminga}
A number of characteristics of the Geminga pulsar have relevance to the likelihood of 
the pulsar, by way of its (past) associated SNR, being important in terms of detected 
cosmic rays, specifically positrons. They can be listed, as follows. \\
1. Geminga is a radio-quiet pulsar and as such, it is less likely than its SNR to have
produced positrons, in view of radio signals being generated by electrons. \\
2. The Geminga pulsar is the second strongest gamma ray source in the sky, a fact known
 since the early SAS II measurements. \\
3. For pulsars of period $P$ bracketing that of Geminga, $P \approx 0.2$ to $0.3 s$,
Geminga has the third highest \.{P} out of 42 in this time window, ie its energy loss
 rate is very high. \\
4. A relevant point that can be made, which may or may not concern Geminga, concerns 
'unusual' types of SN. Some very high mass stars may give rise to 'pair-instability SN'
 (PISN) in which electron-positron pairs play an important role in energy transfer. An 
example is SN 2007bi, in which 3 solar masses of radioactive $^{56}$Ni, a prominent 
positron emitter, were emitted \cite{Gal-Y}. It is possible that the energetic pulsar 
Geminga came from a massive rapidly rotating progenitor star and that the ensuing SNR 
was of the PISN type.

All the above suggest that the progenitor SNR was, like the pulsar itself, energetic 
and thus a proficient accelerator of CR.  
\section{Comparison with the results of other workers}
As mentioned in \S1, there have been many publications explaining the positron upturn 
but, as will be pointed out later, we believe that our work has unique features. 
Kavanaka (2012) has given a useful summary of previous work in the field \cite{Kawan} 
and this can be briefly mentioned. The mechanisms considered can be listed, as follows.
\\
a) Nearby pulsars (~eg \cite{Heyl}~). \\
b) Microquasars (~eg \cite{Heinz}~). \\
c) Dark matter annihilation/decay (~eg \cite{Grass}~). \\
d) SNR with equal numbers of electrons and positrons injected by way of hadronic or
electromagnetic interactions inside the remnant (~eg \cite{Berez} and \cite{Merts}~).\\

Our model differs from those above in a number of ways: \\
(1) It considers that the positrons are generated by the radioactive decay of the SN 
ejecta nuclei, rather than as secondaries - together with the electrons - produced 
within the SNR. \\
(2) It uses anomalous diffusion for the propagation of CR from the SNR to the Earth. 
In fact, the authors of \cite{Volk} have considered such diffusion but in a general way
 without making a specific identification of the source. 

The net result with respect to comparison with the results of other workers is that no 
others appear to have carried out a similar analysis.

Our argument about a pulsar source for the extra positrons is that the expected energy 
spectrum would have the wrong shape.

The case in favour of a SNR is that the expected spectrum is of the correct shape and 
the implied SNR (~the progenitor of the Geminga pulsar~) is not disallowed by its 
non-observation after such a long time. Clearly, the latter is a necessary, but not 
sufficient, condition. Geminga is a strong gamma ray emitter and, as such, suggests 
that its SNR was similarly powerful. Of course, this feature supports a pulsar 
origin, too.

The proposed test, by way of searching for an equivalent upturn in the electron 
spectrum, could confirm the existence of 'new' electrons from a local source and, if 
the excess electrons from  a local source can be identified it should be possibility to
 distinguish between a pulsar and an SNR as the source. Another, very different test 
may eventually appear when measurements of the antiproton to proton ratio are 
available. Antiprotons are not generated in SN ejecta but can come from secondaries in 
the SNR from 'p-p' interactions in the ISM (~eg \cite{Blas2}~). A value of the present 
work is that it indicates that a 'source' at the distance and of the age of Geminga is 
favoured. 

\section{Conclusions}
We argue that the cosmic ray positron excess observed in ATIC-2, Fermi LAT, PAMELA, 
 HESS and recently in the precision AMS-02 experiment can be attributed to the 
production in a local and relatively old supernova remnant. Using the 
prediction of our model of cosmic ray acceleration in SNR we estimate that the SNR 
responsible for the observed positron excess was located between 250 and 320pc from 
the Sun and 170-380 kyear old. The most probable candidate for such a source is the SNR
which contained the Geminga pulsar.   

{\bf Acknowledgements}

The authors are grateful to the Kohn Foundation for financial support.

\end{document}